\documentclass[a4paper,pra,aps,epsfig,twocolumn,showpacs,preprintnumbers,superscriptaddress,float,amsmath,amssymb]{revtex4}

\usepackage{graphicx}
\usepackage{dcolumn}
\usepackage{bm}
\usepackage{physics}
\usepackage{multirow}

\usepackage{color}
\usepackage[normalem]{ulem}

\usepackage{graphicx}

\begin{document}

\title{Adaptive variational preparation of the Fermi-Hubbard eigenstates}

\author{Gaurav Gyawali}
\affiliation{Laboratory of Atomic and Solid State Physics,
Cornell University, Ithaca, NY, 14853}
\author{Michael J Lawler}
\affiliation{Laboratory of Atomic and Solid State Physics,
Cornell University, Ithaca, NY, 14853}
\affiliation{Department of Physics, Applied Physics and Astronomy, Binghamton University, Binghamton, New York 13902}

\begin{abstract}
Approximating the ground states of strongly interacting electron systems in quantum chemistry and condensed matter physics is expected to be one of the earliest applications of quantum computers. In this paper, we prepare highly accurate ground states of the Fermi-Hubbard model for small grids up to 6 sites (12 qubits) by using an interpretable, adaptive variational quantum eigensolver(VQE) called ADAPT-VQE \cite{Grimsley19}. In contrast with non-adaptive VQE, this algorithm builds a system-specific ansatz by adding an optimal gate built from one-body or two-body fermionic operators at each step.
We show this adaptive method outperforms the non-adaptive counterpart in terms of fewer variational parameters, short gate depth, and scaling with the system size. The fidelity and energy of the prepared state appear to improve asympotically with ansatz depth.We also demonstrate the application of adaptive variational methods by preparing excited states and Green functions using a proposed ADAPT-SSVQE algorithm. Lower depth, asymptotic convergence, noise tolerance of a variational approach\cite{VQE14,McClean2016,Sharma_2020} and a highly controllable, system specific ansatz make the adaptive variational methods particularly well-suited for NISQ devices.



\end{abstract}
\maketitle

\section{Introduction}
Finding the ground state of a many-body system is an important problem in condensed matter physics. There are numerous ways of doing so on a classical computer including exact diagonalization, density functional theory(DFT)\cite{DFT} and density matrix renormalization group (DMRG) \cite{DMRG}. However, all of these methods suffer from a rapid scaling of computational resources with the system size for strongly correlated electrons.  Quantum computing has emerged as a promising platform to mitigate the rapid scaling of classical resources because the Hilbert space of the device also grows exponentially. The current hype of quantum computers owes much to the discovery of sub-exponential time quantum algorithms such as Shor's factoring algorithm and quantum phase estimation for classically hard problems. \cite{Shor1997,Mike&Ike}. It is therefore important to study the performance of new quantum algorithms and assess their use in potential breakthrough applications.


Although significant strides have been made in the direction of simulating quantum systems using quantum computers, as envisioned by Feynman\cite{Feynman1982}, we are still in the NISQ era \cite{Preskill2018}. Noisy quantum devices with around $100$ qubits have been available lately, and scientific efforts are already in place to improve the noise tolerance and increase number of qubits \cite{IBM,Google}. A hybrid quantum-classical algorithm called the variational quantum eigensolver(VQE) emerged in the field of quantum chemistry to harness the power of NISQ devices by approximating the ground states of interacting Hamiltonians\cite{VQE14,McClean2016}. VQE requires optimization of a large number of variational parameters but is quite robust to noise and results in a low-depth circuit, which makes it ideal for the near term noisy devices with few qubits.

An algorithm called Adaptive Derivative-Assembled Pseudo-Trotter ansatz VQE (ADAPT-VQE) has been proposed as a quasi-optimal ansatz for a desired level of accuracy \cite{Grimsley19,Zhang2021}. The ansatz is systematically grown by adding optimal fermionic gates one or more at a time such that the maximal amount of correlation energy is recovered at each step. This algorithm generates an ansatz with a small number of parameters and shallow depth, and was tested for various small molecules on quantum hardware. In this paper, we extend the application of ADAPT-VQE algorithm to condensed matter physics by using it to prepare the ground as well as excited states of a strongly correlated electron system called the Fermi-Hubbard model \cite{Hubbard1963}. 

The Hamiltonian of the Fermi-Hubbard model is defined as
\begin{equation} \label{eqn:Fermi-Hubbard}
    H=-t \sum_{<i,j>, \sigma}\left(c_{i \sigma}^{\dagger} c_{j \sigma}+c_{j \sigma}^{\dagger} c_{i \sigma}\right)+U \sum_{i} n_{i \uparrow} n_{i \downarrow},
\end{equation}
where $c_{i \sigma}^{\dagger}$,$c_{i \sigma}$ are fermionic creation and annihilation operators corresponding to  site $i$ and spin $\sigma$; $n_{i,\sigma} = c_{i \sigma}^{\dagger}c_{i \sigma}$, and $\langle.\rangle$ indicates nearest neighbors. $t$ and $U$ are called the hopping and on-site repulsion parameters respectively. Finding the ground state of a Fermi-Hubbard model with uniform $U$ but arbitrary hoppings $(t_{ij}$) has been shown to be in the QMA-complete complexity class \cite{BGorman2021}. Although it implies that the worse-case scenario of the problem is difficult to solve even with a quantum computer, there can be many interesting practical instances that can be solved efficiently \cite{Cao2019}. We focus on a relatively simple yet practical case in which the hopping and interaction parameters are uniform. 

In addition to ground states, we can also use adaptive methods to prepare excited states. This can be done by choosing a variational circuit and applying subspace search VQE(SSVQE).\cite{Endo2020,Nakanishi2019}. The SSVQE algorithm utilizes the conservation of orthogonality under unitary transformation i.e. for a superposition of orthogonal input states, it outputs another superposition of orthogonal states. The circuit is optimized in such a way that these orthogonal output states are the desired excited states. Combining this algorithm with adaptive variational circuits then produces ADAPT-SSVQE, an adaptive method for generating excited states. The quality of these excited states are then demonstrated with a computation of the associated zero-temperature Green's function via the Lehmann representation.

Our results show that ADAPT-VQE algorithm is highly successful at preparing low-depth circuits to approximate ground states of the Fermi-Hubbard model for small grids with high fidelity. We also demonstrate the application of the adaptive variational method to prepare the excited states using ADAPT-SSVQE and thus calculate the Green's function. We show that ADAPT-VQE outperforms state of the art non-adaptive ansatz in terms of performance with the system size, number of parameters, and gate depth. We also observe asymptotic improvement in fidelity as well as energy with each added gate.



\section{ADAPT-VQE algorithm}
\begin{figure}
    \begin{center}
        \includegraphics[scale=0.50]{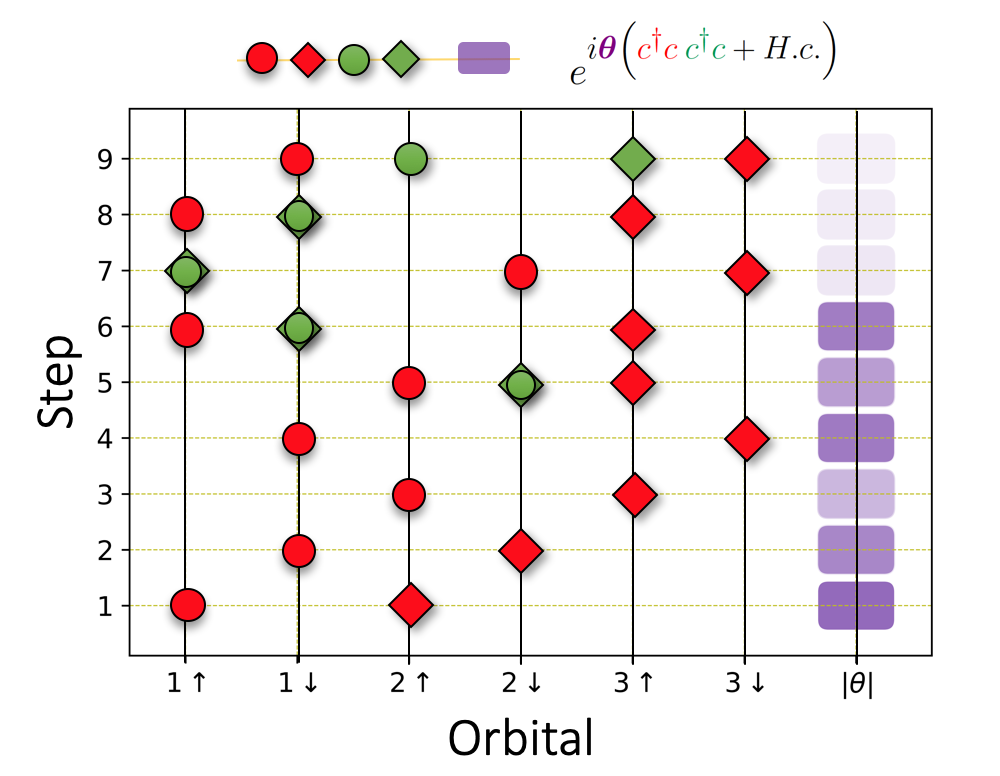}
    \caption{ Pictorial representation of an ADAPT-VQE circuit that prepares the ground state of a 3 $\times$ 1 Fermi-Hubbard model with $t=1$ and $U=3$. Creation and annihilation operators are denoted by a circle and a diamond respectively. A set of diamond and a circle represent a one-body hopping term ($c^\dagger_{i,\sigma} c_{j,\sigma}$) whereas a set of four shapes represent a two-body term ($c^\dagger_{i,\sigma} c_{j,\sigma} c^\dagger_{k,\sigma'} c_{l,\sigma'} $ ). Two-body terms with $k=l$ are the correlated hopping terms for which a circle and a diamond overlap. Transparency of the purple box indicates the absolute value of the coefficient($\theta$)  for that step with respect to the maximum coefficient.}
    \label{fig:3x1_operators}
    \end{center}
\end{figure}

The ADAPT-VQE ansatz is built one operator at a time as opposed to the non-adaptive ansatzs such as Hamiltonian variational ansatz and number preserving ansatz in which the operators are fixed, and only optimization performed is for the variational parameters \cite{HVA2015, Cade2019}. Each operator in this ansatz is chosen from a pool of one-body and two-body Hamiltonians $\hat{A}_m = \{ \hat{H}_{ij}, \hat{H}_{ijkl} \}$, where
\begin{align}
    \hat{H}_{ij} &= c^{\dagger}_{i} c_{j} + H.c. \\
	\hat{H}_{ijkl} &= c^{\dagger}_{i} c^{\dagger}_{j} c_{k} c_{l} + H.c.,
\end{align}
where $i,j,k,l$ are qubit labels. These Hamiltonians are chosen such that they are spin conserving. The adaptive ansatz is applied on top of the initial-state preparation circuit, which is chosen to be either a product state or a Slater determinant. At the $n^{th}$ step of the ansatz, the energy gradient for each of the operator ($\hat{A}_m$) i.e.
\begin{equation}
    \frac{\partial E^{(n)}}{\partial \theta_{m}}=\left\langle\psi^{(n)}\left|\left[\hat{H}, \hat{A}_{m}\right]\right| \psi^{(n)}\right\rangle
\end{equation}
is calculated. Here, $\ket{\psi^{(n)}}$ is the optimized wavefunction at the $n^{th}$ step. The wavefunction at $(n+1)^{th}$ step is then given by
\begin{equation}
    \ket{\psi^{(n+1)}} = e^{i\theta_{n+1}\hat{A}_{n+1}} \ket{\psi^{(n)}},
\end{equation}
where $\hat{A}_{n+1}$ is the $\hat{A}_{m}$ with highest energy gradient. It is important that we optimize the accumulated parameter set $\vec{\theta}^{(n+1)} = \{ \vec{\theta}^{(n)}, \theta_{n+1}$ \} at each step rather than using $\vec{\theta}^{(n)}$ from the previous step and just optimizing $\theta_{n+1}$. This process is repeated until stopping criteria are met i.e.
$(E^{(n)} - E^{(n-1)}) < \epsilon$ or $\abs{\pdv{E^{(n)}}{\theta_i}} < \delta$. See Ref. \cite{Grimsley19} for a more detailed description of this algorithm. The variational algorithm was implemented using OpenFermion-FQE library available in Python \cite{fqe21}. Mapping from fermionic to qubit operators was done via Jordan-Wigner transform \cite{JW28} as described in Appendix \ref{sec:JW_transform}. We used L-BFGS-B algorithm available in SciPy library to perform all optimizations\cite{SciPy}.

Formulation of the ADAPT-VQE algorithm in terms of fermionic operators makes it convenient to interpret an ansatz. The $n^{th}$ step of an ansatz involving operator $\hat{A}_n$ and variational parameter $\theta_n$ can be thought of as the time evolution of the system under $\hat{A}_n$ for time $-\theta_n$ i.e. $e^{-i \hat{A}_n (-\theta_n)}$. At each step only the most optimal gate is selected and so earlier steps are more important than later ones. A pictorial representation of the ADAPT-VQE optimized ansatz for a $3 \times 1$ Fermi-Hubbard model is shown in Fig. \ref{fig:3x1_operators}. Here, the first four operators are one-body/hopping gates between qubits corresponding to different sites \emph{and same spin}. They create a state close to the ground state for the non-interacting part of the Hamiltonian with the same spin symmetry. The fifth operator introduces two-body terms that apply the corrections due to interaction. Of these corrections, majority are three qubit gates i.e. correlated hopping. These gates appear more often when the interaction is weak, and the system size is small.  We have observed that the ground state of a Hamiltonian with no interaction (not shown in the figure) can be prepared only by using the hopping gates as expected from Thouless theorem \cite{Thouless1960}. Introducing the interaction therefore adds three-qubit and four qubit gates and their relative numbers is determined by the system size and interaction strength. Hence the ADAPT-VQE ansatz tells us the spin symmetry, importance of interactions, and may even give us a sense of how electrons move around in the material via correlated hopping processes.

\section{Green's function}
A fundamental object of interest in many-body physics is the Green's function. It contains many important dynamic and static properties of a many-body system including the quasi-particle dispersion\cite{Coleman2015}. Throughout this paper, we'll focus on Green's function at zero temperature.  Consider a  fermionic system described by Hamiltonian $H$. The retarded-time Green's function for this system is defined as follows \cite{Mahan1990,Coleman2015}:
\begin{equation}
    G(t)=-i \Theta(t)\left\langle c_{a}(t) c_{b}^{\dagger}(0)+c_{b}^{\dagger}(0) c_{a}(t)\right\rangle,
    \label{eqn:Greens}
\end{equation}
where $c^{\dagger}_a$, $c_a$ are creation and annihilation operators for a fermionic mode $a$; for instance $a = (k,\sigma)$, with $k$ being the momentum, and $\sigma$ the spin. $\Theta(t)$ is the Heaviside step function, and $c_a(t)$ is a Heisenberg operator defined as $c_a(t) = e^{iHt} c_a(0) e^{-iHt}$. The expectation value is calculated with respect to the ground state $\ket{G}$ of $H$. Taking the Fourier transform of Eq. \ref{eqn:Greens} gives us the propagator $G(k,\omega)$. The imaginary part of the propagator gives us information about the spectrum of the many-body system, and thus known as the spectral function: 
\begin{equation}
    A(k,\omega)=-\frac{1}{\pi} \operatorname{Im} G(k,\omega).
\end{equation}
We can write the retarded Greens function in Lehmann representation, which relates the definition in Eq. \ref{eqn:Greens} to the eigenspectrum of $H$, as follows:

\begin{eqnarray}
 && G(k,\omega)= \nonumber\\
 &&\sum_{n}\left(\frac{\left|\left\langle E_{n}\left|c^\dagger_k \right| \mathrm{G}\right\rangle\right|^{2}}{\omega+E_{\mathrm{G}}-E_{n}+i \nu}+\frac{\left|\left\langle E_{n}\left|c_{k}\right| \mathrm{G}\right\rangle\right|^{2}}{\omega-E_{\mathrm{G}}+E_{n}+i \nu}\right), 
\end{eqnarray}
where $E_n$ is the $n^{th}$ excited energy of $H$ corresponding to the eigenstate $\ket{E_n}$, and $E_G$ is the ground state energy.

\subsection*{Green's function by ADAPT-SSVQE}
The SSVQE algorithm can prepare a desired number of excited states on a quantum computer with a single variational circuit by using the orthogonality conservation property of a unitary transformation\cite{Nakanishi2019,Endo2020}. The algorithm results in a variational quantum circuit $U(\vec{\theta})$ which acts on a set of orthonormal states  $\{ \ket{\psi_j}_{j=0}^{K-1} \}$ to approximate the lowest $K$ eigenstates of a Hamiltonian $H$ i.e. $\ket{E_j} \approx U(\vec{\theta}) \ket{\psi_j}, 0<j<K$. In this paper, we focus on an adaptive construction of  $U(\vec{\theta})$ by using ADAPT-VQE with the following cost function:
\begin{equation}
    \mathcal{C}(\vec{\theta})=\sum_{j=0}^{K-1} w_{j}\left\langle\psi_{j}\left|U(\vec{\theta})^{\dagger} H U(\vec{\theta})\right| \psi_{j}\right\rangle = \sum_{j=0}^{K-1} w_{j} E_j,
\end{equation}
where $\vec{\theta}$ represents the variational parameters, and $\mathrm{w}_0 > \cdots \mathrm{w}_{K-1}>0$ are the weights which are chosen to ensure that the eigenenergies are in ascending order i.e. $E_0 < E_1 \cdots E_{K-1}$. By minimizing this cost function, we tune the variational parameters such that $ U(\vec{\theta}) \ket{\psi_j}$ is approximately $\ket{E_j}$, the $j^{th}$ eigenstate of $H$. Once we have the approximate eigenstates  $\ket{E_j}$ and eigenenergies $E_j$ for both particle and hole excitations, we can obtain the transition amplitudes of creation and annihilation operators, and thus use Lehmann representation to compute the propagator. A detailed description on how to calculate the transition amplitudes is given in Ref. \cite{Endo2020}. Since the number of excited states increases exponentially with the system size, it quickly become impossible to keep track of the full spectrum, so we can focus on the first $K$ excited states and carry out the Green's function calculation, which will still give us valuable information about the low energy excitations.

\section{Results and discussion}
\subsection{Ground state results}

\begin{figure}[]
    \begin{center}
        \includegraphics[scale=0.5]{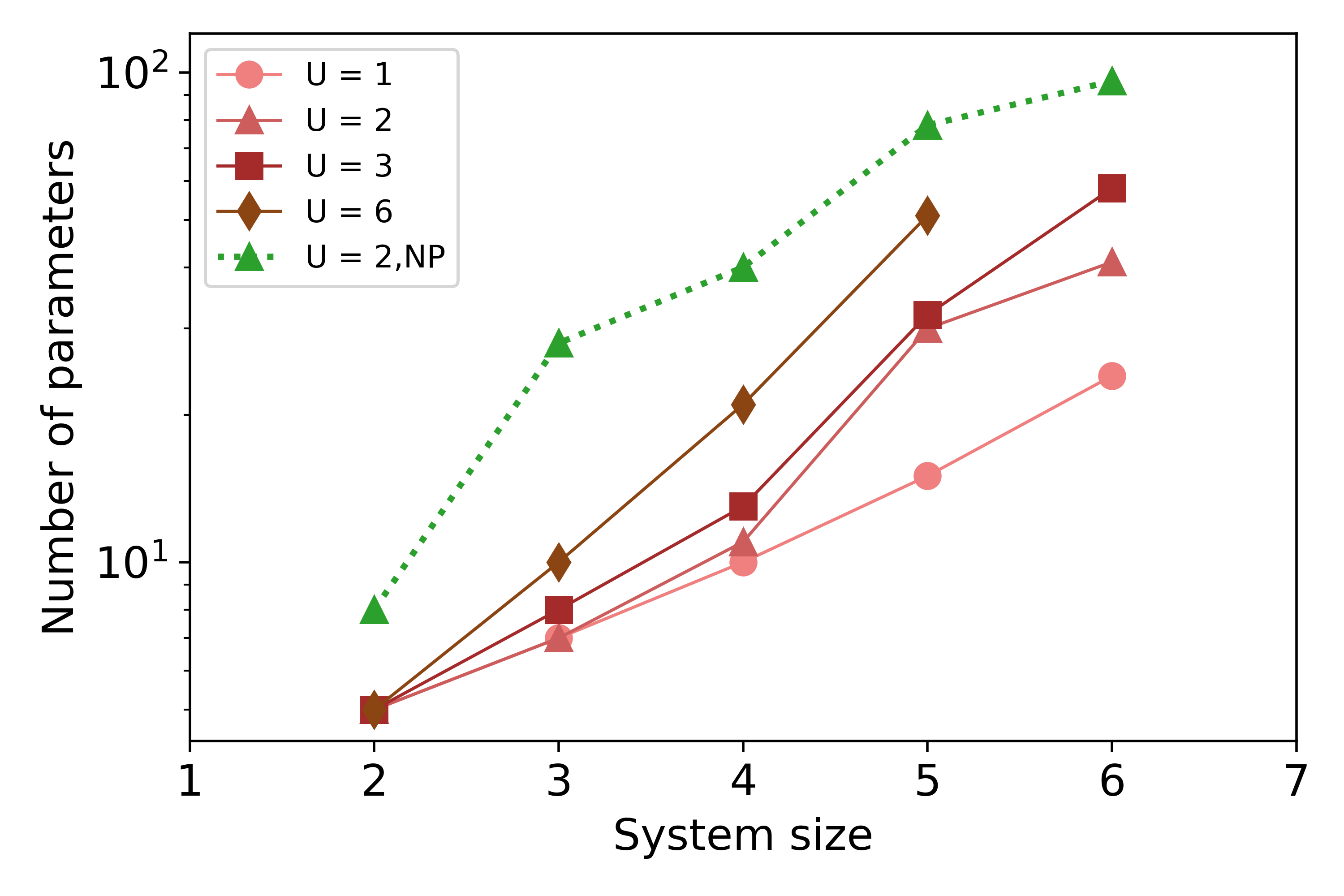}
    \caption{Scaling of the number of parameters required to achieve $0.99$ fidelity with the system size for 1-dimensional Fermi-Hubbard model. Solid lines indicate ADAPT-VQE algorithm whereas dashed line is for non-adaptive algorithm using a number preserving (NP) ansatz, taken from Ref.\cite{Cade2019}.}
    \label{fig:Depth_vs_size}
    \end{center}
\end{figure}

\begin{figure}[]
    \begin{center}
        \includegraphics[scale=0.5]{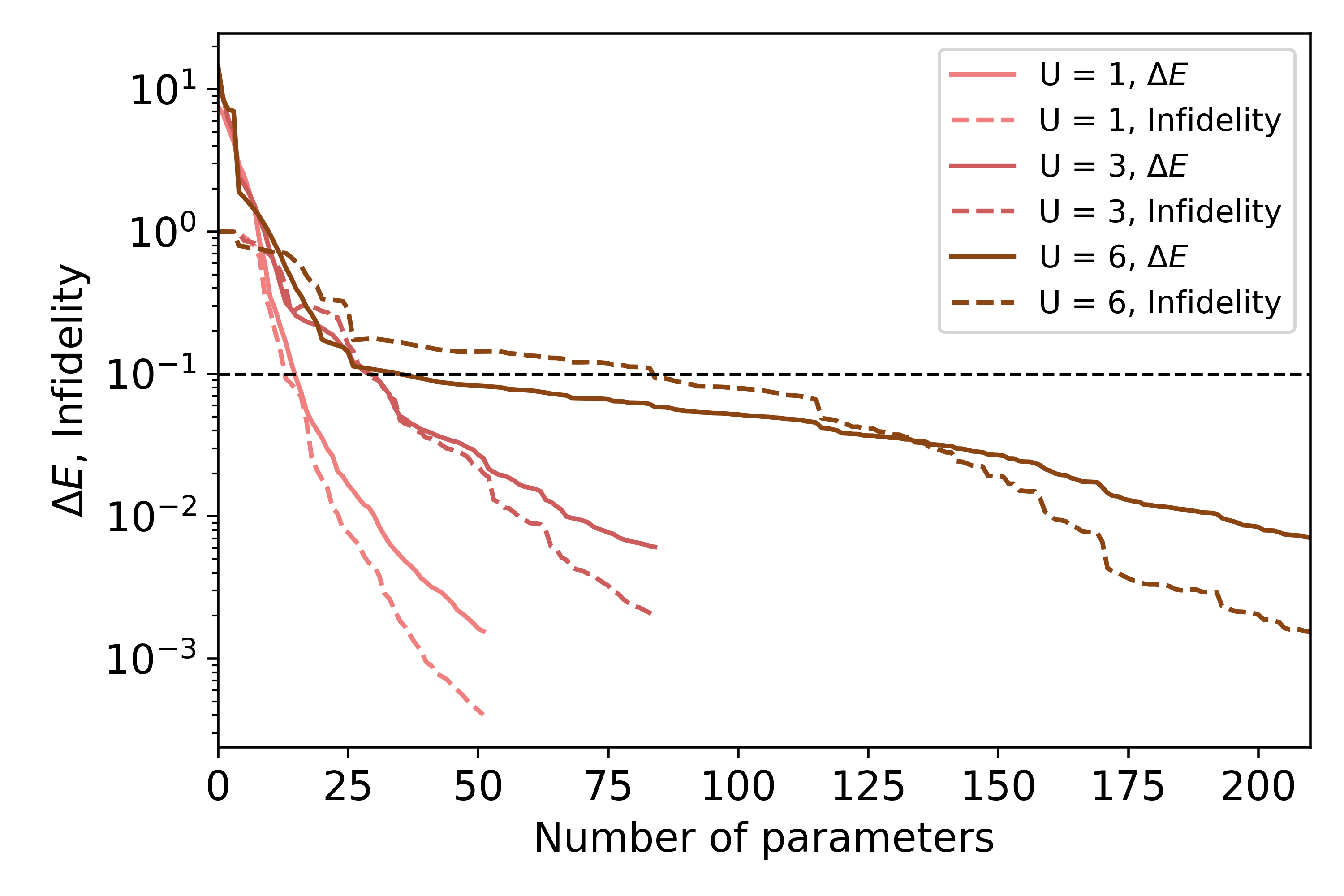}
    \caption{Scaling of infidelity(1-fidelity) and error in energy ($\Delta E = E -E_{exact} $) with the number of steps for a 3 $\times$ 2 grid. Dashed black line indicates 0.99 fidelity. Notice that the accuracy increases asymptotically with the number of gates. }
    \label{fig:3x2_fid_and_dE}
    \end{center}
\end{figure}

\begin{table}[]
    \centering
    \begin{ruledtabular}
\begin{tabular}{cccccc}
Grid &  U &  $\text{E}_\text{exact}$ &  Depth &  Fidelity &                 $\Delta E$ \\
\hline
    \multirow{3}{*}{$2 \times 1$} &  1 &         -2.56 &      4 &      1.00 &    $3.80 \times 10^{-12}$ \\
                                  &  3 &         -4.00 &      4 &      1.00 &    $1.03 \times 10^{-11}$ \\
                                  &  6 &         -6.61 &      4 &      1.00 &    $1.15 \times 10^{-11}$ \\
\hline
  \multirow{3}{*}{$3 \times 1$}  &  1 &         -3.51 &      9 &      1.00 &   $1.17 \times 10^{-09}$ \\
                                &  3 &         -5.12 &      9 &      1.00 &   $2.81 \times 10^{-09}$ \\
                                &  6 &         -7.85 &     10 &      1.00 &   $1.52 \times 10^{-10}$ \\
\hline
 \multirow{ 3}{*}{$4 \times 1$}    &  1 &         -5.58 &     19 &      1.00 &   $2.90 \times 10^{-04}$ \\
                                 &  3 &         -8.35 &     24 &      1.00 &   $1.17 \times 10^{-04}$ \\
                                 &  6 &        -13.43 &     29 &      1.00 &   $1.42 \times 10^{-03}$ \\
\hline
 \multirow{3 }{*}{$2 \times 2$}    &  1 &         -5.34 &     35 &      1.00 &   $2.87 \times 10^{-08}$ \\
                                 &  3 &         -8.42 &     32 &      1.00 &   $6.52 \times 10^{-05}$ \\
                                 &  6 &        -13.63 &     32 &      1.00 &   $2.19 \times 10^{-05}$ \\
\hline
 \multirow{3}{*}{$5 \times 1$}    &  1 &         -6.73 &     32 &      1.00 &   $7.88 \times 10^{-04}$ \\
                              &  3 &         -9.74 &     53 &      1.00 &   $1.50 \times 10^{-03}$ \\
                              &  6 &        -14.99 &     75 &      1.00 &   $7.15 \times 10^{-04}$ \\
\hline
 \multirow{3 }{*}{$6 \times 1$}    &  1 &         -8.63 &     51 &      1.00 &   $1.56 \times 10^{-03}$ \\
                                 &  3 &        -12.72 &     84 &      1.00 &   $6.09 \times 10^{-03}$ \\
                                 &  6 &        -20.27 &     79 &      0.98 &   $2.19 \times 10^{-02}$ \\
\hline
 \multirow{3}{*}{$3 \times 2$}    &  1 &         -9.28 &     52 &      1.00 &   $1.12 \times 10^{-03}$ \\
                                &  3 &        -13.28 &     80 &      0.99 &   $2.93 \times 10^{-02}$ \\
                                &  6 &        -20.73 &     71 &      0.88 &   $6.74 \times 10^{-02}$ \\
\end{tabular}
\end{ruledtabular}
    \label{tab:fid_and_dE}
    \caption{Final depth, fidelity and error in energy ($\Delta E = E -E_{exact} $) reached for the adaptive ansatz optimized variationally using the L-BFGS-B method available in SciPy. Stopping criteria were  $(E^{(n)} - E^{(n-1)}) < 10^{-3}$ and $\abs{\pdv{E^{(n)}}{\theta_i}} < 10^{-4}$, where the partial derivative is taken with respect to the $i^{th}$ parameter at depth $n$.}
\end{table}

We successfully used the ADAPT-VQE algorithm to create quantum circuits that produce high fidelity ground states of the Fermi-Hubbard model with various system sizes up to 6 sites (12 qubits). We explored these systems for interaction strength($U$) up to 6, and the method works well for the range tested. Table \ref{tab:fid_and_dE} summarizes the final depth of the variational circuit as well as achieved fidelity and the error in energy ($\Delta E = E -E_{exact} $) for the following stopping criteria: $(E^{(n)} - E^{(n-1)}) < 10^{-3}$ and $\abs{\pdv{E^{(n)}}{\theta_i}} < 10^{-4}$. Depth here refers to the number of fermionic gates constituting the ansatz, which is also equal to the number of variational parameters. The scaling of the ansatz depth required to achieve a fidelity of $0.99$ with the system size for a 1-dimensional Fermi-Hubbard model is shown in Fig.  \ref{fig:Depth_vs_size}.  

In Ref. \cite{Cade2019}, the authors constructed ground states for the Fermi-Hubbard model by using non-adaptive methods such as the Hamiltonian variational ansatz and the number-preserving ansatz to reach 0.99 fidelity. Our circuits use significantly less parameters to reach the same fidelity. For instance, they report that 160 parameters are required to reach 0.99 fidelity for 3 $\times$ 2 grid, $U=2$, and we obtain a better fidelity with just 62 parameters. In Fig. \ref{fig:Depth_vs_size}, we compare the scaling of the number of parameters between number preserving ansatz from Ref.\cite{Cade2019} and ADAPT-VQE ansatz for $U=2$, and it can be seen that ADAPT-VQE scales better. Although for the system sizes studied, the scaling appears to be exponential, the number preserving ansatz scales sub-exponentially for larger system sizes{\cite{Cade2019}}, so the adaptive ansatz can be expected to exhibit a similar behavior since it outperforms number-preserving ansatz in every single study we performed.


We also found that fidelity as well as error in energy improve consistently with the number of gates, and the wavefunction becomes exact in the limit of infinite depth. Such scaling for the largest grid that we studied is shown in Fig. \ref{fig:3x2_fid_and_dE}. As expected, the number of steps required to produce the ground state for a given fidelity increases with the strength of the interaction ($U$). Convergence was not an issue for any of the systems that we studied although the runtime tended to increase with increased system size, interaction, and desired accuracy. For those problems in which estimating the ground state energy of a strongly interacting system is the main goal, reasonable estimates can be obtained with just a few steps. The first few ($\sim$ number of sites) gates in our optimized ansatz are predominantly hopping terms involving only two qubits, which further adds to the shallowness of the circuit. In total, it appears from these results that arbitrary accuracy can be obtained at the cost of increased gate depth and number of parameters.

The ADAPT-VQE method is particularly well suited for NISQ devices compared to the non-adaptive  Hamiltonian inspired variational ansatz \cite{Cade2019}. In our adaptive method, a step involves adding one (or a specified number) operator to the ansatz from a given pool. Non-adaptive methods usually require addition of all the interaction terms present in the Hamiltonian at each step. The order and type of gates used in the non-adaptive methods is not optimized at each step, so they are not as flexible as the adaptive methods in terms of when we want to terminate circuit. Since the NISQ devices are noisy, having a deeper circuit will add more noise than add to the accuracy of ground state. The stopping criteria $\epsilon$ makes ADAPT-VQE aware of when exactly adding a gate stops being advantageous. Since there is always going to be some number of layers before the stopping criteria are met, progress is always made towards some understanding of the ground state due to the interpretability of our ansatz.


It would seem that because we treat the one-body and two-body fermionic operators as our fundamental gates rather than focus on their implementation in terms of single-qubit and entangling gates, our approach would be inefficient on a real NISQ device. For example, the number of such gates required to perform these operations on quantum hardware would greatly depend on the specific architecture i.e. connectivity between the qubits and natural gates available. But for real-device implementation, the ADAPT-VQE algorithm allows the cost of implementing these fermionic operators to be incorporated into the cost function. So we can optimize the ansatz not only for the accuracy of the result but also for the depth of the resulting circuit.

\subsubsection*{Choice of initial state}
\label{subsection:initial_state}
\begin{figure}[]
    \begin{center}
        \includegraphics[scale=0.32]{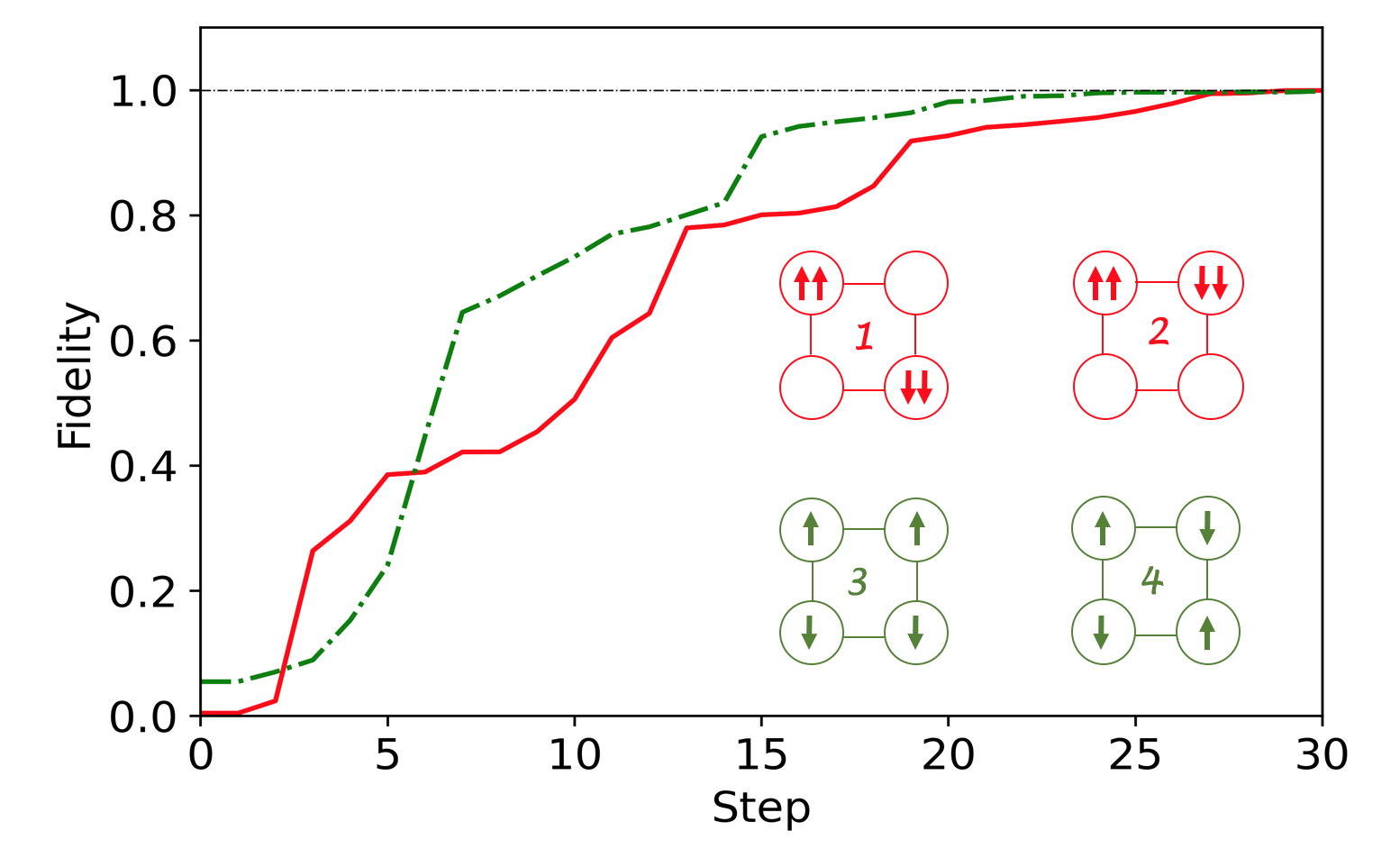}
    \caption{Comparision of fidelity vs. starting configuration for 2 $\times$ 2. grid with $U=3$. Red line corresponds to configurations 1 and 2 for which two of the sites are doubly occupied and remaining two are unoccupied whereas green line corresponds to configurations 3 and 4 for which all sites are singly occupied.  }
    \label{fig:2x2_fid}
    \end{center}
\end{figure}
We performed variational optimization of adaptive ansatz using both non-interacting ground state (i.e. Slater determinants) and product state as the initial wavefunction, and found that the product state always produces the ground state with desired accuracy with less number of total gates. Since the operator pool is formed by number-preserving operators, it is important to start in the subspace with desired occupancy, which can be done by applying X-gates on the number of qubits equal to the occupation number. For all the grids that we studied, the best choice of the product state was to evenly spread out the fermions in the grid because of being close to the half filling. Although with enough steps, all starting distributions can result in a ground state with arbitrary accuracy, spread-out initial states reach the desired accuracy faster. One such example for 2 $\times$ 2 grid with $U=3$ is shown in Fig. \ref{fig:2x2_fid}. We note that the fidelity-scaling is dependent only on the number of singly-occupied vs. doubly-occupied sites and not on their permutations. We thus note any prior understanding of the ground state electron density is rewarded in the form of lower depth.


\subsection{Excited states and Green's function}
\begin{figure}[h!]
    \begin{center}
        \includegraphics[scale=0.39]{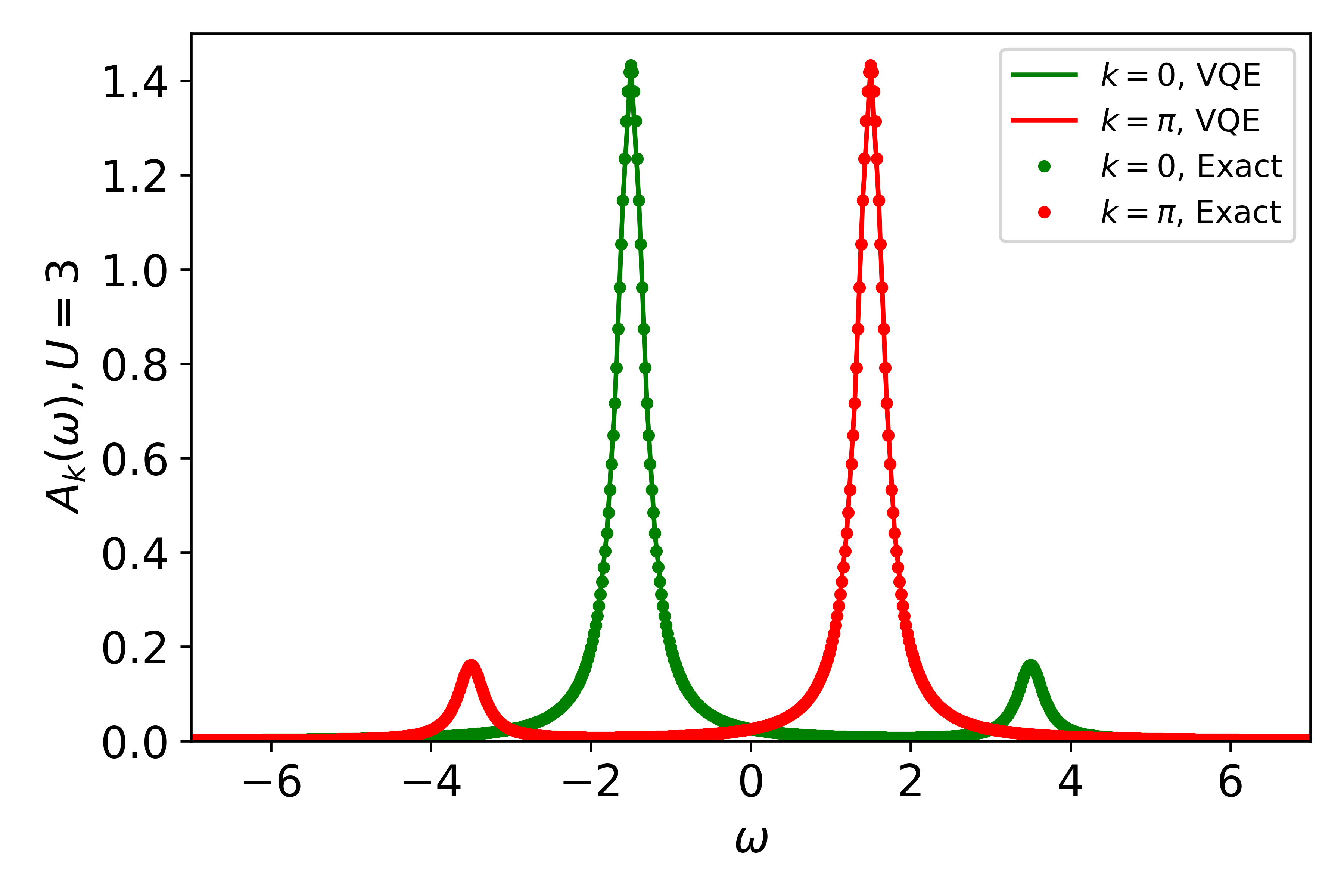}
        \includegraphics[scale=0.39]{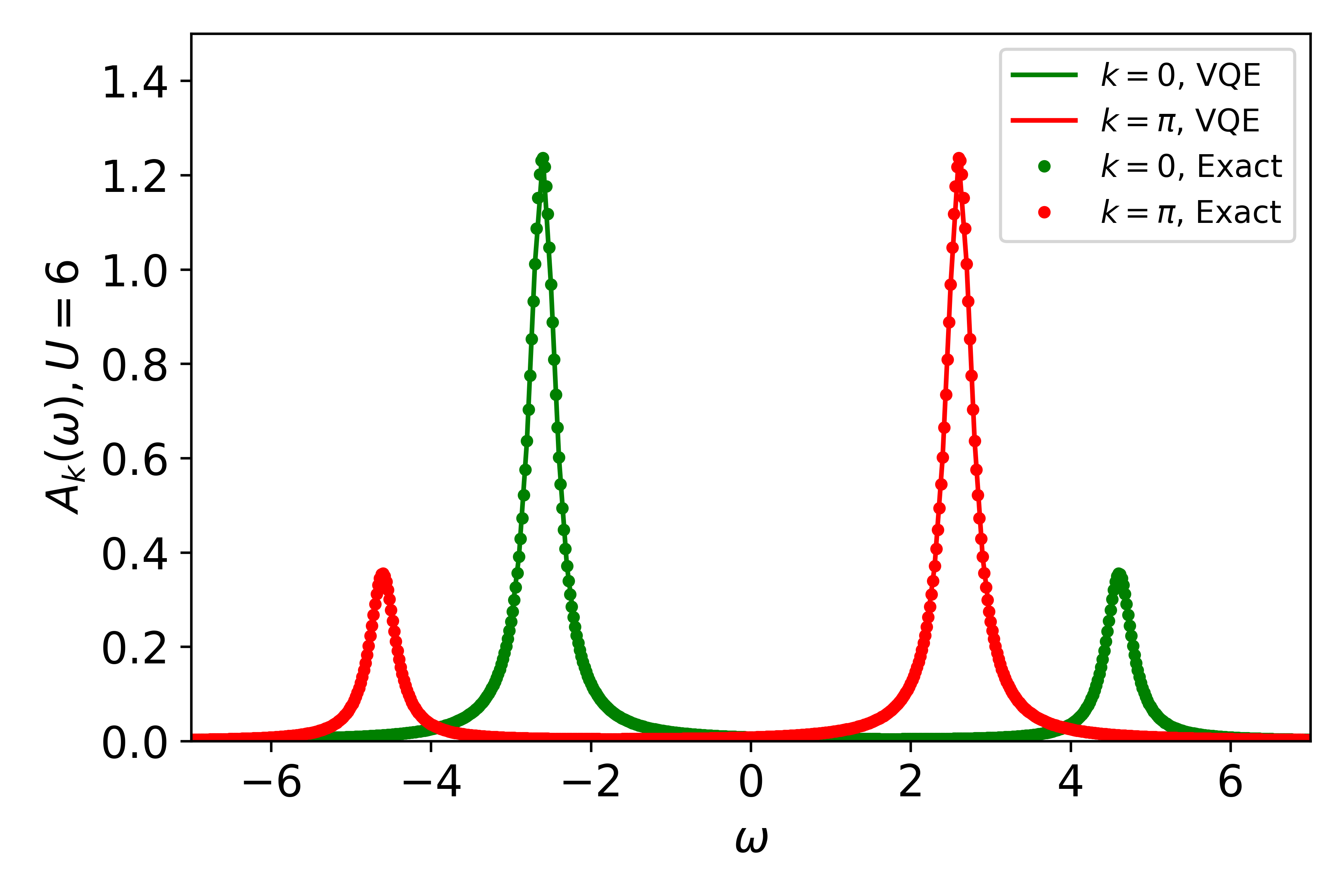}
    \caption{Spectral functions calculated for Fermi-Hubbard model on a 2 $\times $ 1 grid at half filling for $U=3$ (top) and $U=6$ (bottom) by using ADAPT-SSVQE method.}
    \label{fig:green}
    \end{center}
\end{figure}
We also show that ADAPT-SSVQE (i.e. ADAPT-VQE + SSVQE) is effective at preparing the excited states of the Fermi-Hubbard model, and hence helps us predict the spectral properties. We used the SSVQE method with different weights to compute the ground state for the simple case of 2 $\times$ 1 grid with $U=3$ and $U=6$ \cite{Endo2020}. The ADAPT-SSVQE-computed spectral functions for these specific cases, along with the ones obtained by exact diagonalization, are shown in Fig. \ref{fig:green}. Ref. \cite{Endo2020} reports using 48 parameters to calculate the excited states for calculating the Green's function using a non-adaptive ansatz whereas the adaptive algorithm was able to do so with just 5 parameters. The number of parameters caps the number of excited states that can be obtained using an ansatz. The effectiveness of the adaptive variational algorithm to calculate Green's function for this rather small system points towards a promising future in terms of simulating condensed matter systems using quantum computers. 

We note that the excitation spectrum of the Hamiltonian is related to the dynamics, hence the ADAPT-VQE algorithm also has potential application to variational quantum simulation(VQS)\cite{Li_2017,Endo2020}.

\section{Conclusion}
In this paper, we demonstrate the application of adaptive variational methods (ADAPT-VQE and proposed ADAPT-SSVQE) to prepare the ground as well as excited states of the Fermi-Hubbard model of strongly correlated electron systems on small grids. In contrast with the non-adaptive methods, a system-specific ansatz is built by adding one or more fermionic operators at a time, resulting in fewer variational parameters and gate depth. The energy and fidelity of the prepared states improve asymptotically with the addition of each operator. Lower depth, consistent convergence, and highly controllable system-specific ansatz make this method suitable to studying strongly interacting problems in condensed matter and quantum chemistry on NISQ devices.

\section*{Acknowledgement}
The authors would like to thank Nicholas Mayhall for his helpful suggestions on the manuscript.

\newpage
\bibliographystyle{unsrt}
\bibliography{main_pra.bib}

\appendix
\section{Jordan-Wigner transform} \label{sec:JW_transform}
We used the Jordan-Wigner transform\cite{JW28} to map the fermionic creation and annihilation operators into Pauli operators (i.e. tensor product of Pauli matrices). The Jordan-Wigner transform is used to map a $n-$fermionic system to a $n-$qubit system as follows:
\begin{eqnarray}
    a_{k} &&\mapsto\left(\prod_{j=0}^{k-1} Z_{j}\right) \sigma_{k}^{-} \nonumber\\ 
    &&=\frac{1}{2}\left(Z_{0}\otimes \cdots \otimes Z_{k-1} \otimes\left(X_{k}-i Y_{k}\right)\right),
\end{eqnarray}
where $X_{j}, Y_{j}, Z_{j}$ are the Pauli spin operators, and $\sigma_{j}^{+/-}$ are the ladder operators acting on qubit $j$. By virtue of this transform, the qubit operators, which follow commutation relations,  are mapped to creation/annihilation operators, which follow anticommutation relations.

\section{Choice of Fillings}
\begin{table}[h]
    \centering
    \begin{ruledtabular}
    \begin{tabular}{cc} 
     Number of electrons & Grid \\ 
     \hline
     2 & 2$\times$ 1, 3 $\times$ 1   \\ 
     4 & 4 $\times$ 1, 2 $\times$ 2, 5 $\times$ 1 \\ 
     6 & 6 $\times$ 1, 3 $\times$ 2 \\
    \end{tabular}
    \end{ruledtabular}
    \label{tab:fillings}
    \caption{Number of electrons/fillings and the corresponding grid sizes chosen such that the Fermi-Hubbard Hamiltonian has non-degenerate ground states}
\end{table}

The fillings were chosen such that the Fermi-Hubbard Hamiltonian has a non-degenerate ground state. Degeneracy gives rise to ambiguity in the choice of ground state to use for calculating the fidelity.

\end{document}